\documentclass[twocolumn]{aastex701}

\usepackage{graphicx} 
\usepackage{amsmath}
\usepackage{hyperref}
\usepackage{savesym}
\usepackage{siunitx}
\usepackage{lineno}   

\newcommand{\Msun}{M_{\odot}}
\newcommand{\Lsun}{L_{\odot}}

\begin{document}
\linenumbers 

\title{WISE Photometry of Galaxies within 10 Mpc}

\author[0000-0002-9291-1981]{R. Brent Tully}
\affiliation{Institute for Astronomy, University of Hawaii. 2680 Woodlawn Drive, Honolulu, HI 96822, USA}
\email{tully@ifa.hawaii.edu}

\author[0000-0002-5514-3354]{Ehsan Kourkchi}
\affiliation{Institute for Astronomy, University of Hawaii. 2680 Woodlawn Drive, Honolulu, HI 96822, USA}
\email{ekourkch@gmail.com}

\author[0000-0002-0466-1119]{James D. Neill}
\affiliation{Division of Physics, Mathematics and Astronomy, California Institute of Technology, 1200 E. California Blvd.. 91125, CA, USA}
\email{neill@astro.caltech.edu}

\begin{abstract}
    For photometry of galaxies in the infrared that extend to very low surface brightnesses, it is necessary to observe with telescopes in space.
    WISE and NEOWISE provide sensitive all-sky access to mid-infrared flux from galaxies at $3.4\mu$m (W1) and $4.6\mu$m (W2) bands. 
    This study is complemented by the availability of accurate Tip of the Red Giant Branch distances for a large fraction of the targets.
    In this work, photometry is assembled on a complete volume limited sample of galaxies between one and ten Mpc brighter than absolute magnitude $-13$ in the W1 band.
    Stellar masses are inferred from W1 fluxes and measured distances.
    While most of the galaxies are dwarfs, most of the stellar mass is concentrated in a small number of giants.
    
\end{abstract}

\section{Introduction}
\label{sec:intro}

This article describes a program to obtain photometry in WISE W1 band for galaxies suspected to lie within 10~Mpc.
The motivation is to compile a reasonably complete inventory of the mass in stars within this limit (with an exclusion about the plane of the Milky Way).
Such an inventory can be acquired for high surface brightness galaxies from 2MASS, the Two Micron All Sky Survey \citep{2000AJ....119.2498J}.
However, 2MASS did not have the sensitivity for accurate photometry of low surface brightness galaxies of the sort that make up the vast majority of systems in a volume limited domain \citep{2008AJ....136.1866K, 2009ApJ...703..517D}.
The infrared flux from these faint targets that best correlates with stellar mass \citep{2014ApJ...782...90C, 2014MNRAS.445..899C} can only be  acquired in space.

Nearby galaxies have been observed in the infrared with both Spitzer Space Telescope and the Wide-field Infrared Survey Explorer (WISE).
There have been two legacy surveys with Spitzer: SINGS, the Spitzer Infrared Nearby Galaxy Survey \citep{2003PASP..115..928K} and LVL, the Spitzer Local Volume Legacy survey \citep{2009ApJ...703..517D}.
The former observed 75 galaxies within 30~Mpc, 32 of these within 10~Mpc.
The latter observed 258 galaxies at $\vert b \vert > 20^{\circ}$ within 11~Mpc to complement 11HUGS, the 11~Mpc H${\alpha}$ and Ultraviolet Galaxy Survey \citep{2008ApJS..178..247K, 2011ApJS..192....6L}.
The SINGS sample was also given attention with WISE \citep{2017ApJ...850...68C, 2025ApJ...979...18C}.  See also the S$^4$G Spitzer observations of a sample extending to 40~Mpc \citep{2018ApJS..234...18B, 2025A&A...697A..38S}.

The LVL survey was extensive, but not complete.
The current program complements ongoing imaging with the Hubble Space Telescope (HST) that resolves brighter stars in nearby galaxies and permits an estimate of distances from the magnitudes of stars at the TRGB, the tip of the red giant branch \citep{2006AJ....131.1361K, 2007ApJ...661..815R, 2009ApJS..183...67D, 2019ApJ...885..141B, 2021AJ....162...80A}.
With either Advanced Camera for Surveys (ACS) or Wide Field Camera-3 (WFC3) on HST, the reach in a single orbit for a reliable distance estimate is $\sim 10$~Mpc.
There are now 490 galaxies with TRGB distance estimates in EDD, the Extragalactic Distance Database \citep{2009AJ....138..323T} in the catalog CMDs/TRGB.  Of these, 408 lie beyond the Local Group ($d>1$~Mpc) and within the HST one orbit detection limit of 10~Mpc.

These galaxies with TRGB distances provide the backbone of the current sample.
Additional galaxies are included that are suspected to also lie within 10~Mpc.
Candidates are continually being discovered.
However, most are faint; indeed, fainter than $M_B=-13$.
The aspiration of this program is to have a complete inventory of galaxies with distances $1<d<10$~Mpc and brighter than a WISE W1 magnitude limit that approximates $M_B<-13$.
With that intent, the sample errs to include fainter and more distant galaxies.
Also, although obviously galaxies are lost due to obscuration at low Galactic latitudes, there is no explicit exclusion zone.
The full sample is comprised of 567 galaxies.

\section{Photometry with WISE}
\label{sec:phot}

Whereas observations with the Spitzer Space Telescope involved  target pointing and potentially rastering of larger candidates given a $4^{\prime}$ field of view, WISE enables extractions from an all-sky survey.
Hence, any object of interest can be given attention although survey acquisitions are terminated.
The initial survey \citep{2010AJ....140.1868W} in four bands at $3.4 \mu$m (W1), $4.6 \mu$m (W2), $12 \mu$m (W3), and $23 \mu$m (W4) continued until the exhaustion of cryogens.
After a pause, the observations successfully continued at the two shorter passbands, W1 and W2, as NEOWISE \citep{2014ApJ...792...30M}.

Here, full attention is given to the W1 band, with successively lesser completion at the longer wavelength bands.
Mosaics at the coordinates of targets are constructed using the ICORE, Image Co-addition with Optical Resolution Enhancement, software \citep{2009ASPC..411...67M, 2013ascl.soft02010M}.
Image cutouts are drizzled to give combined images with output scale $1.0^{\prime\prime}$~pixel$^{-1}$.
The resolution in the W1 band is $6^{\prime\prime}$.

Photometry is carried out with the GLGA package \citep{2009ApJ...707.1449N}.
Fluxes are measured in successive elliptical apertures with fixed centers, shapes, and orientations, stepping from the center of the galaxy to the edge.
Foreground stars and other contamination are masked and there is an accounting for their areal coverage.
In recognition that faint foreground stars superimposed on the galaxy cannot be reliably identified, stars at a similar level are not removed from the adjacent sky field in establishing the surface photometry growth curve. 

Default centers, axial ratios, and position angles are taken from HyperLEDA \citep{2003A&A...412...45P}.
Each could be modified upon inspection.
The most sensitive parameter is the setting of the sky background level.
The software provides two graphic representations of particular value.
One gives the accumulated flux (on the logarithmic magnitude scale) as a function of the major axis radius.
The other records the run of surface brightness with major axis radius.
If the flux growth curve continues to rise at large radius then the sky level must be set too low such that sky is contributing to the galaxy count; if the growth curve falls then the sky level is set too high.
With the second test, the run of surface brightness in magnitudes~arcsec$^{-2}$ typically decays as a power law with radius.
A break upward at large radius suggests the sky is contributing and a break downward suggests galaxy counts are given to the sky.
With manipulation, the sky contribution is optimized to satisfy these tests.
Typically, the signal from the galaxy is strong down to surface brightness $\sim27$~mag as$^{-2}$.
Extremely low surface brightness dwarfs can be lost below this threshold set by background but such galaxies have absolute magnitudes much fainter than our completeness limit discussed in Section~\ref{sec:properties}. 

Two magnitudes are recorded, an aperture magnitude and an asymptotic magnitude.  The aperture magnitude can be extrapolated assuming a continuation of the exponential decay to infinity.
The asymptotic magnitude is a direct measure of the total magnitude to the limit that the growth curve goes flat.

Small corrections are made following \citet{2014ApJ...792..129N}.
There is an aperture correction because the WISE photometric calibration is conducted with point sources with a fixed aperture that misses extended flux that would be included in photometry of extended sources; the corrections $A_a^{W1}=-0.034$~mag and $A_a^{W2}=-0.041$~mag are applied.
In addition,there is a correction for Milky Way extinction \citep{2011ApJ...737..103S}, generally small in the mid-infrared.
Extinction can also arise internally in the target galaxies \citep{2019ApJ...884...82K}.
It is significant in larger spiral galaxies but is negligible in the dwarf galaxies that comprise the bulk of a volume limited sample.

Results of the photometry are summarized in Table~\ref{table:wise_catalog}.


\begin{table*}

\centering
\caption{10 Mpc WISE W1/W2 photometry catalog, and galaxy structural parameters.}
\label{table:wise_catalog}
\label{tab:big_catalog_split}

\begingroup
\scriptsize
\setlength{\tabcolsep}{3pt}
\renewcommand{\arraystretch}{0.95}

\begin{tabular*}{\textwidth}{@{\extracolsep{\fill}} r l r r r r r c r r r r r}
\hline\hline
PGC & Name & RA & Dec & 1PGC & $V_{LS}$ & $d$ & Flag & $M_B$ & $\log L_K$ & $\log M_\star$ & $M_{W1}$ & $W1$ \\
    &      & (deg) & (deg) &  & (km s$^{-1}$) & (Mpc) &  & (mag) & ($L_\odot$) & ($M_\odot$) & (mag) & (mag) \\
(1)&(2)&(3)&(4)&(5)&(6)&(7)&(8)&(9)&(10)&(11)&(12)&(13)\\
\hline
35      & UGC12894          & 0.09375 & 39.49550 & 35      & 618  & 8.84 & 1 & -13.84 & 7.605 & 7.35 & -13.22 & 16.531 \\
143     & WLM               & 0.49260 & -15.46093& 143     & -31  & 0.96 & 1 & -14.03 & 7.674 & 7.52 & -13.63 & 11.275 \\
5056918 & Andromeda\_XVIII  & 0.56040 & 45.08889 & 5056918 & -41  & 1.28 & 1 & -8.24  & 6.407 & 6.09 & -10.06 & 15.503 \\
388     & ESO409-015        & 1.38300 & -28.09832& 388     & 753  & 8.51 & 1 & -14.59 & 8.092 & 7.65 & -13.95 & 15.698 \\
5059815 & AGC748778         & 1.64295 & 15.51083 & 5059815 & 479  & 6.08 & 1 & -10.02 & 6.361 & 7.42 & -13.39 & 15.541 \\
5056920 & Andromeda\_XX     & 1.87785 & 35.13233 & 2557    & -182 & 0.84 & 1 & -5.62  & 5.281 & 4.82 & -6.88  & 17.749 \\
591     & UGC00064          & 1.93365 & 40.87590 & 591     & 588  & 7.97 & 1 & -13.90 & 7.995 & 7.70 & -14.08 & 15.439 \\
621     & ESO349-031        & 2.05635 & -34.57873& 621     & 211  & 3.14 & 1 & -11.87 & 7.105 & 6.57 & -11.26 & 16.219 \\
701     & NGC0024           & 2.48475 & -24.96383& 701     & 595  & 7.13 & 1 & -17.27 & 9.419 & 9.25 & -17.97 & 11.303 \\
930     & NGC0045           & 3.51645 & -23.18207& 930     & 515  & 6.48 & 1 & -17.75 & 9.295 & 9.38 & -18.29 & 10.765 \\
\nodata \\
\hline
\end{tabular*}

\vspace{8pt}

\begin{tabular*}{\textwidth}{@{\extracolsep{\fill}} r r r r r r r r r r r r c}
\hline\hline
PGC & $W2$ & $A_{W1}$ & $A_{W2}$ & $\mu_{e,W1}$ & $\mu_{e,W2}$ & $\mu_{90,W1}$ & $\mu_{90,W2}$ & $R_{e,W1}$ & $R_{e,W2}$ & $R_{90,W1}$ & $R_{90,W2}$ & QW \\
    & (mag)& (mag) & (mag) & (mag/arcsec$^2$) & (mag/arcsec$^2$) &
(mag/arcsec$^2$) & (mag/arcsec$^2$) & (arcmin) & (arcmin) & (arcmin) & (arcmin) & \\
(1)&(14)&(15)&(16)&(17)&(18)&(19)&(20)&(21)&(22)&(23)&(24)&(25)\\
\hline
35      & 17.791 & 0.021 & 0.016 & 24.99 & 25.66 & 27.58 & 26.20 & 0.33 & 0.27 & 0.66 & 0.38 & 4 \\
143     & 11.730 & 0.007 & 0.006 & 24.32 & 24.93 & 26.68 & 26.62 & 3.19 & 3.43 & 6.63 & 7.15 & 4 \\
5056918 & 16.438 & 0.020 & 0.016 & 23.74 & 24.23 & 25.52 & 26.06 & 0.28 & 0.24 & 0.55 & 0.47 & 3 \\
388     & 15.981 & 0.003 & 0.003 & 23.30 & 23.56 & 25.46 & 25.72 & 0.25 & 0.25 & 0.50 & 0.52 & 5 \\
5059815 & 17.083 & 0.012 & 0.010 & 26.39 & 28.20 & 0.00 & 27.45 & 1.32 & 1.09 & -0.02 & 1.31 & 5 \\
5056920 & 17.686 & 0.011 & 0.009 & 26.73 & 26.46 & 26.47 & 26.59 & 0.47 & 0.48 & 0.58 & 0.59 & 1 \\
591     & 16.076 & 0.015 & 0.012 & 23.56 & 23.99 & 25.90 & 27.01 & 0.30 & 0.28 & 0.75 & 0.68 & 5 \\
621     & 16.937 & 0.002 & 0.002 & 25.57 & 26.79 & 27.06 & 26.72 & 0.42 & 0.42 & 0.78 & 0.91 & 4 \\
701     & 12.021 & 0.004 & 0.003 & 21.45 & 21.91 & 23.97 & 23.98 & 1.03 & 0.95 & 2.49 & 2.15 & 0 \\
930     & 11.298 & 0.004 & 0.003 & 23.57 & 24.34 & 25.64 & 27.49 & 2.19 & 2.34 & 4.55 & 5.53 & 0 \\
\nodata \\
\hline
\end{tabular*}
\vspace{3pt}
\noindent{\scriptsize\textit{Note.} The complete version of this catalog is available at \url{https://edd.ifa.hawaii.edu}.}

\vspace{3pt}
\noindent{\scriptsize\textit{Column descriptions.}
(1) PGC identifier;
(2) galaxy name;
(3) right ascension (deg);
(4) declination (deg);
(5) PGC ID of the dominant galaxy in the group;
(6) Local Sheet velocity \citep{2008ApJ...676..184T} (km s$^{-1}$);
(7) distance (Mpc);
(8) distance flag (1 = TRGB, 0 = not TRGB; from UNGC);
(9) absolute $B$ magnitude from RC3 when available;
(10) $\log_{10}(L_{K}/L_{\odot})$, total $K$-band luminosity;
(11) $\log_{10}(M_{\star}/M_{\odot})$, stellar mass inferred from W1 luminosity as discussed in Section~\ref{sec:properties};
(12) absolute W1 magnitude (AB), reddening corrected, $M_{\odot}=5.91$~mag;
(13) apparent asymptotic W1 magnitude; typical uncertainty 0.05 mag
(14) apparent asymptotic W2 magnitude; typical uncertainty 0.05 mag
(15--16) reddening at W1 and W2 (mag);
(17--20) W1/W2 surface brightness at effective and 90\% radii (mag arcsec$^{-2}$);
(21--24) W1/W2 effective and 90\% radii along the major axis (arcmin);
(25) WISE photometry quality flag; qualitative assessment evaluating disturbance, confusion, surface brightness issues (5 = best, 1 = worst, 0 = uncertain).}
\endgroup
\end{table*}

\section{Evaluation of the WISE W1 Magnitudes}
\label{sec:eval}

\begin{figure}[!]
\centering
\includegraphics[width=1.00\linewidth]{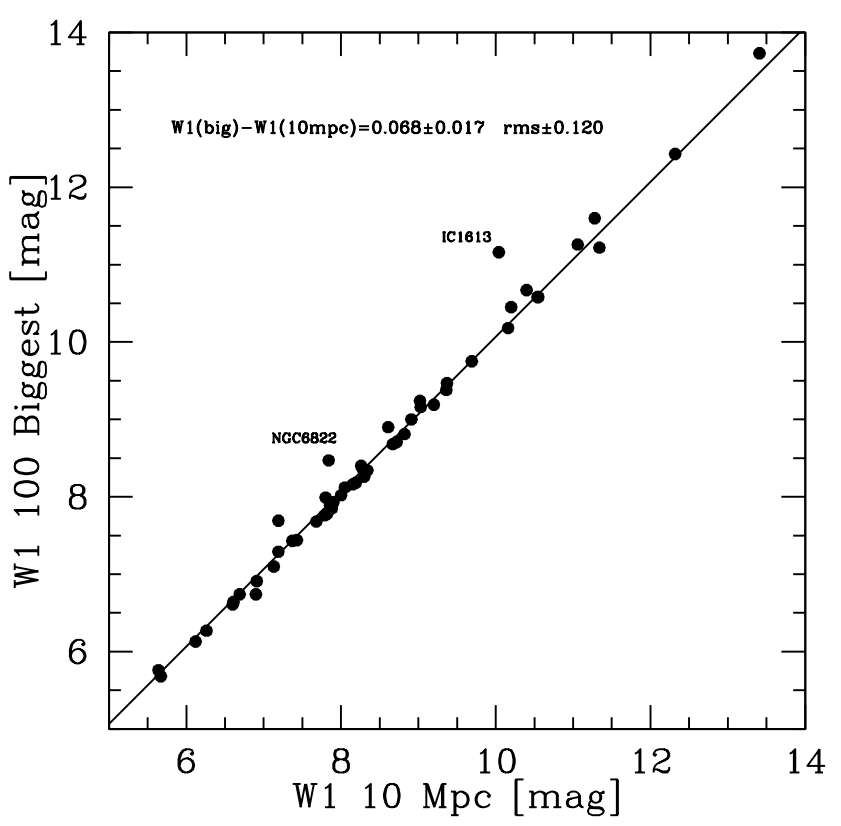}
\caption{WISE W1 magnitudes alternatively from this program (horizontal axis) vs from the WISE 100 largest galaxies program (vertical axis).}
\label{fig:wise2}
\end{figure}

\begin{figure}[]
\centering
\includegraphics[width=1.00\linewidth]{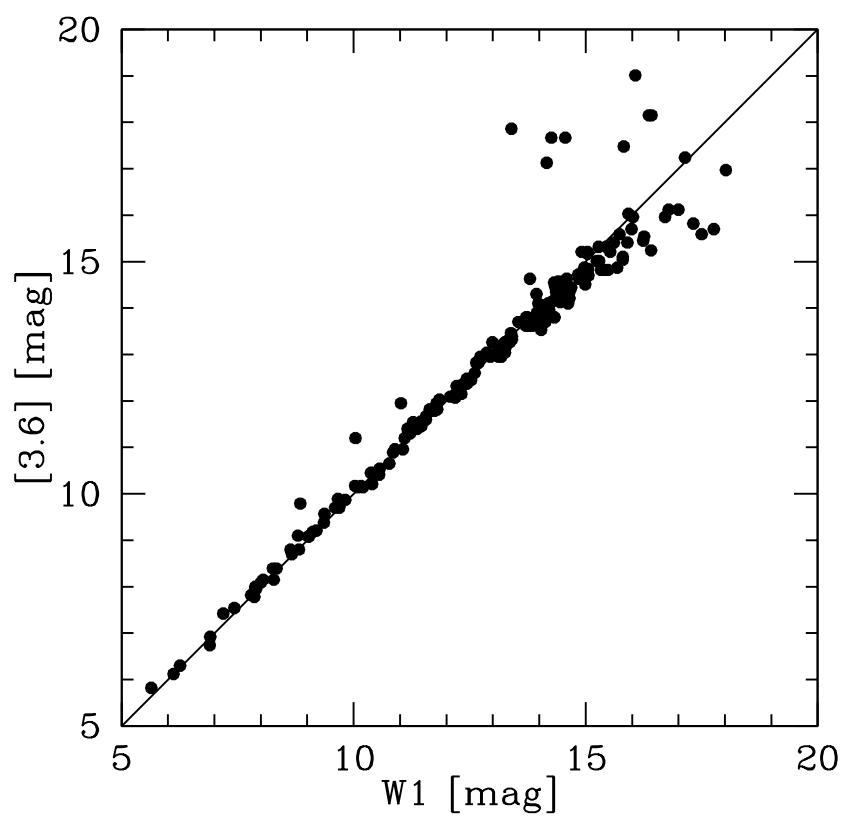}
\caption{WISE W1 v Spitzer [3.6] magnitudes.  The straight line is the one-to-one correspondence; not a fit.}
\label{fig:wise_spitzer}
\end{figure}

\begin{figure}[!]
\centering
\includegraphics[width=1.00\linewidth]{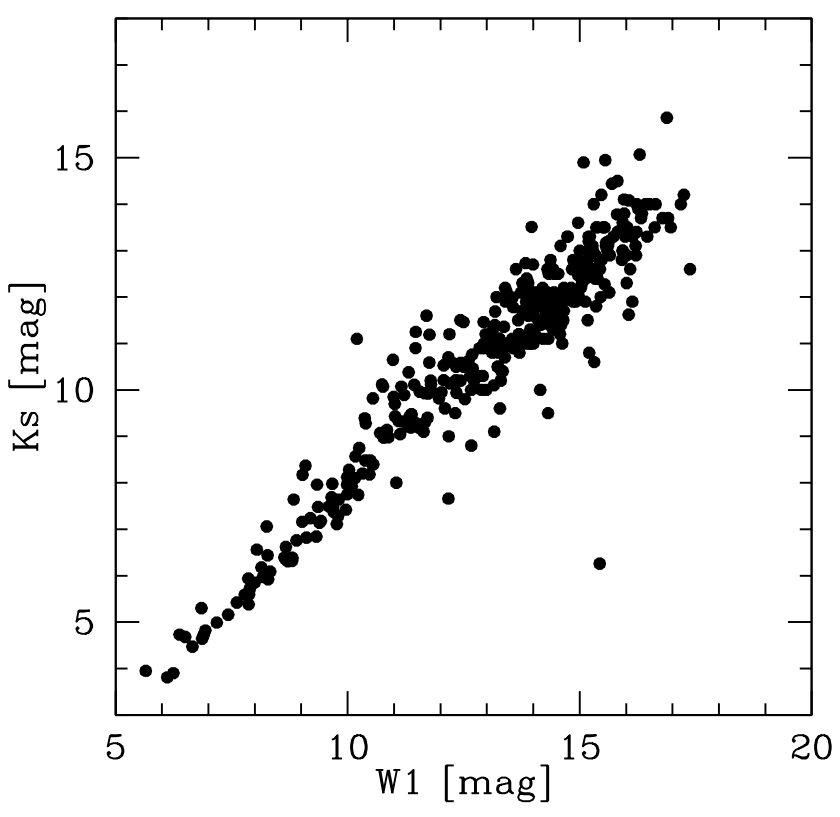}
\caption{WISE W1 v 2MASS Ks magnitudes.}
\label{fig:w9se_2mass}
\end{figure}

\begin{figure}[!]
\centering
\includegraphics[width=1.00\linewidth]{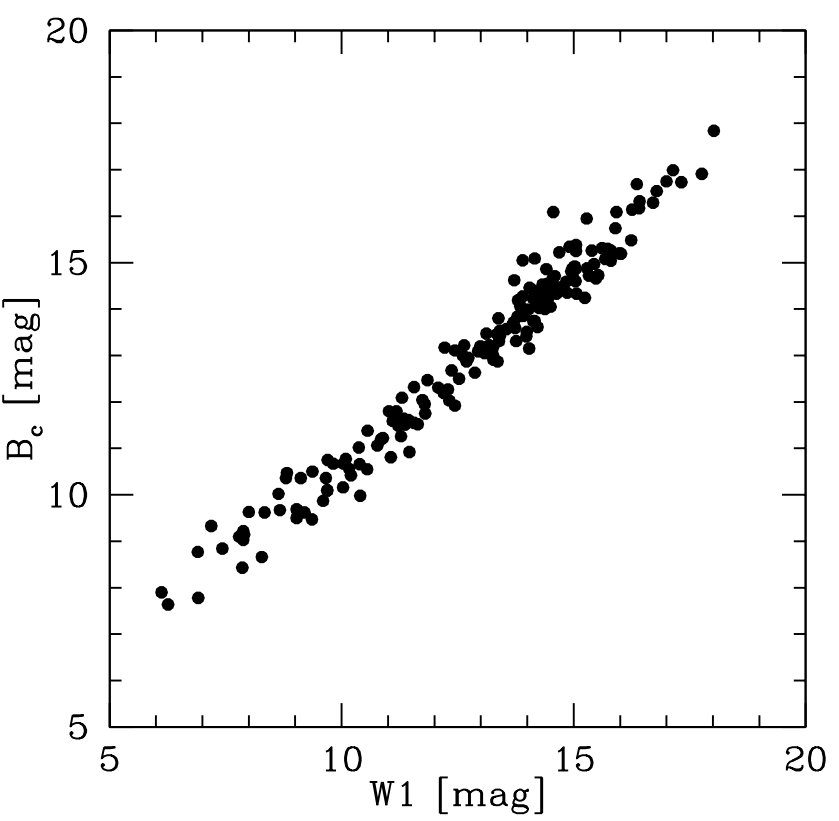}
\includegraphics[width=1.00\linewidth]{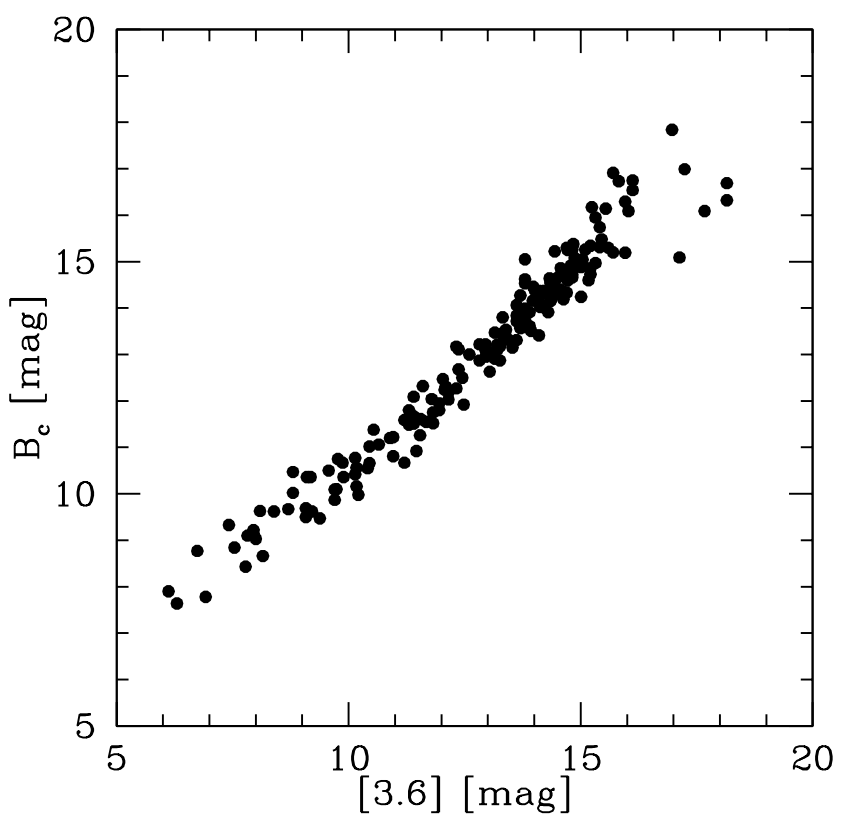}
\caption{Top: B v W1 magnitudes, Bottom: B v [3.6] magnitudes. All on AB scale.}
\label{fig:Bcompare}
\end{figure}

Figure~\ref{fig:wise2} compares WISE W1 magnitudes from this study with W1 magnitudes for 53 galaxies in common from the 100 largest galaxies study of \citet{2019ApJS..245...25J}, while
Figure~\ref{fig:wise_spitzer} compares WISE W1
magnitudes with Spitzer [3.6] aperture magnitudes \citep{2009ApJ...703..517D}. All are on the AB scale.  
The straight lines show the one-to-one correspondence between the alternate samples.  These are not fits, but show that the scales are similar.  In Fig.~\ref{fig:wise2}, discounting discordance with the Local Group low surface brightness galaxies IC\,1613 and NGC\,6822, there remains an offset of 0.068 mag (our magnitudes brighter).  In Fig.~\ref{fig:wise_spitzer}
agreement begins to break down at W1$\sim$[3.6]$\sim$14~mag.  Progressively, W1 magnitudes are fainter than [3.6] magnitudes beyond 14 mag.
Figure~\ref{fig:w9se_2mass} compares WISE W1 magnitudes with 2MASS Ks magnitudes \citep{2000AJ....119.2498J, 2003AJ....125..525J} on the same plotting scale.  The superiority of the space observations for these mostly low surface brightness galaxies is evident.

The LVL survey provides B band magnitudes from a combination of direct UBVR observations or transformed from SDSS ugri photometry \citep{2014MNRAS.445..881C, 2014MNRAS.445..890C}.
Figure~\ref{fig:Bcompare} compares these B magnitudes, corrected for Galactic and internal reddening, with alternatively WISE W1 and Spitzer [3.6] magnitudes.
With the Spitzer comparison there is a noticeable bend at [3.6]~$\sim11.5$ mag.
The B v W1 relation is close to linear, with a slight hint of a bend around W1~$\sim13$~mag.
The sense of the bend in the Spitzer case is consistent with the departures from the fiducial line in Figure~\ref{fig:wise_spitzer}; that is, Spitzer magnitudes brighter than W1 at the faint limit.
It is to be noted that the W1 sample extends to fainter galaxies than the Spitzer sample.  The Figure~\ref{fig:Bcompare} comparisons reach B~$\sim17$~mag, with [3.6] magnitudes scattering beyond [3.6]~$\sim15$~mag.
The W1 sample in its entirety reaches $\sim 19$~mag.
In the ensuing discussion,WISE W1 will be given preference over Spitzer [3.6] mags.
Uncertainties are small to at least W1$\sim 17.5$ corresponding to $M_{W1}\sim-12.5$ at the outer edge of the sample.
Photometry for galaxies fainter than $M_{W1}=-13$ is recorded in Table~\ref{table:wise_catalog} but will not be given further attention.

\section{Properties of the W1 Sample}
\label{sec:properties}

\begin{figure}[!]
\centering
\includegraphics[width=1.00\linewidth]{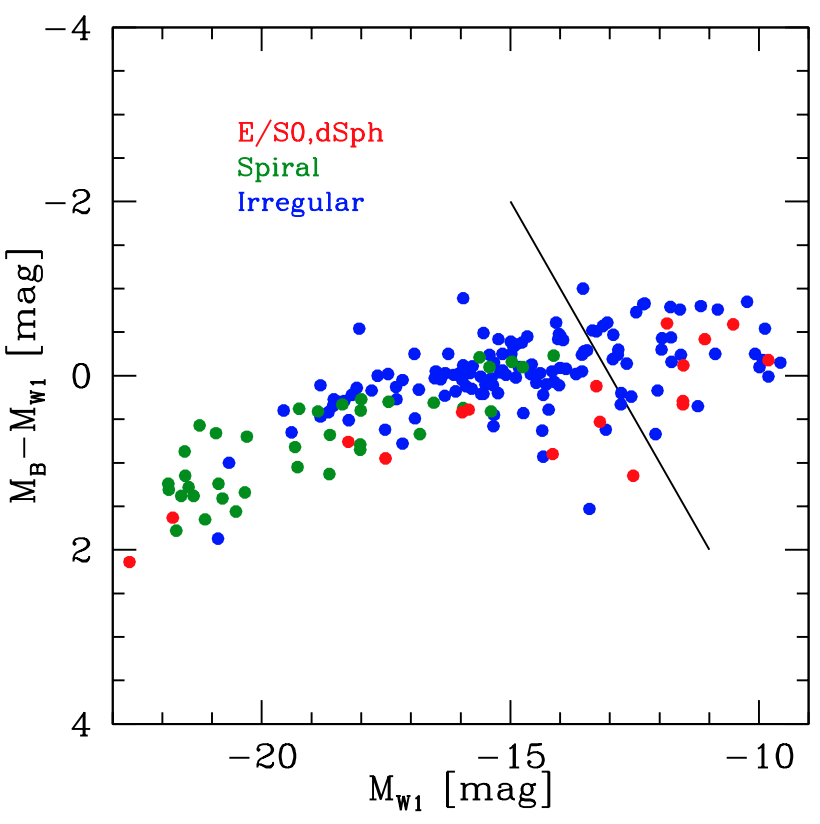}
\caption{B$-$W1 color as a function of absolute W1 magnitude. Galaxy types are identified by colors: gas-poor in red, spirals in green, and dwarf irregulars in blue. Straight line: $M_B = -13$.}
\label{fig:absWB}
\end{figure}

\begin{figure}[!]
\centering
\includegraphics[width=1.00\linewidth]{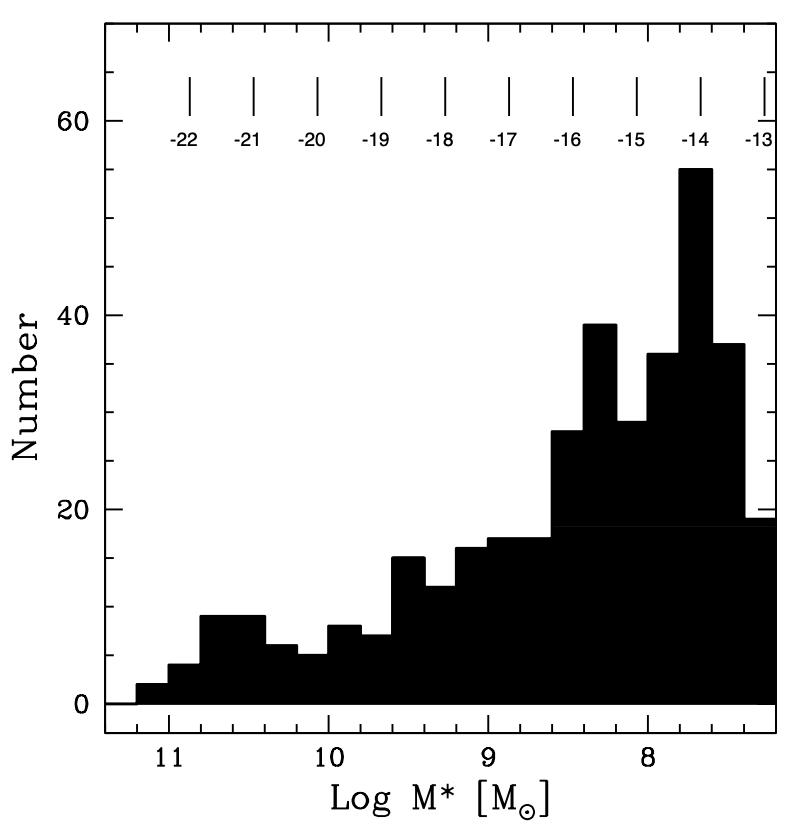}
\includegraphics[width=1.00\linewidth]{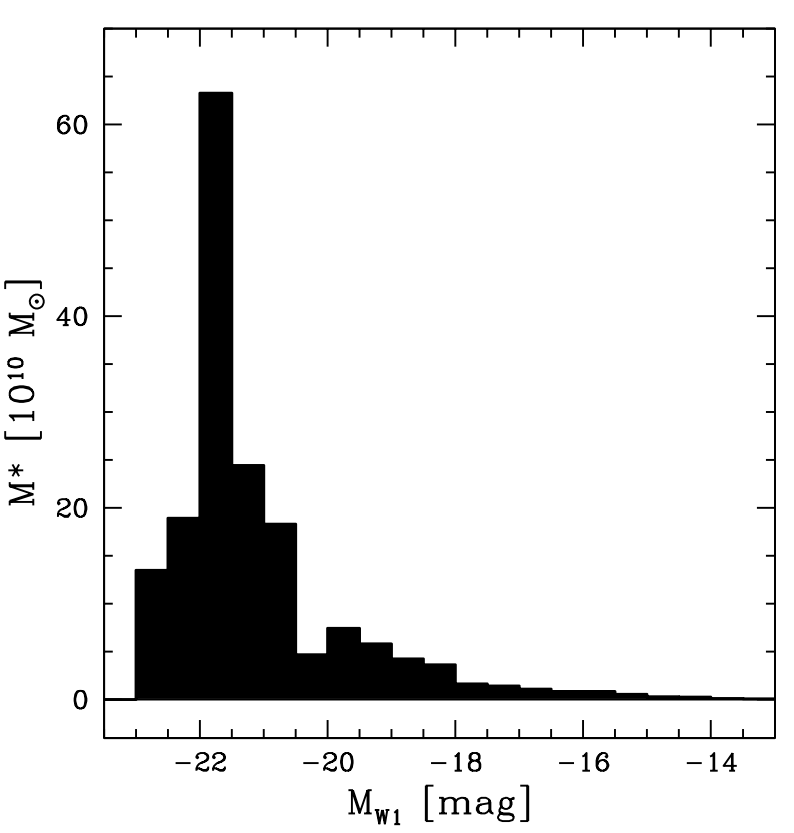}
\caption{Top: histogram of the 1 to 10~Mpc, $M_{W1}<-13$~mag sample in increments of log stellar mass, with corresponding $M_{W1}$ intervals indicated across the top. Bottom: histogram of stellar mass for the same sample, in units of $10^{10}~\Msun$, in intervals of $M_{W1}$~mag.}
\label{fig:2masshist}
\end{figure}

The top panel of Figure~\ref{fig:Bcompare} is recast in Figure~\ref{fig:absWB} as an optical to mid-infrared color versus absolute W1 magnitude.
Now, in addition to the availability of both $B$ and W1 magnitudes,  distances are required and eighty-five percent of distances contributing to Figure~\ref{fig:absWB} are of high accuracy from TRGB measurements.
Mixed quality distance estimates for the remainder are from UNGC, the {\it Updated Nearby Galaxy Catalog} \citep{2013AJ....145..101K}.
The sample selection cut at $M_B=-13$ translates to a cut in $M_{W1}$ shown by the straight line in  Fig.~\ref{fig:absWB}.  It is seen that for the main color trend $M_{W1} \simeq M_B$ at the cut line.

The trend toward bluer colors at fainter magnitudes \citep{2004ApJ...613..898T} results from a mix of age (star formation rates) and metallicity effects.
The mean stellar populations in gas-poor systems are older, reflected in the E/S0/dSph symbols lying to the red at given absolute magnitudes.\footnote{Morphological types are from the Third Reference Catalog \citep{1991trcb.book.....D} except in a small number of cases of galaxies erroneously typed dSph with CMD manifesting star formation.}
Decreasing metallicity with lower galaxy mass shifts stellar envelopes on the red giant branch to hotter colors, driving the trend to bluer colors at lower luminosities. 

Luminosities in the mid infrared W1 band arise from the Rayleigh-Jeans tail of the galaxy spectral energy distribution dominated by evolved stars.  \citet{2014ApJ...782...90C} provide a recipe for the translation from W1 luminosities to stellar mass, $M^{\ast}$, that involves $W1-W2$ colors.  In the current nearby galaxies sample there is scatter, not taken to be significant, around the average value $W1-W2=-0.60$~AB mag.
With the correspondence +0.04 Vega mag, \citet{2014ApJ...782...90C} find $\Upsilon_{\ast}^{W1} = \rm{log}_{10} M^{\ast}/L_{W1} = 0.54~\Msun/\Lsun$ for resolved sources.  
This value is close to that given in a summary of literature results of $\Upsilon_{\ast}^{W1} = 0.5~\Msun/\Lsun$ \citep{2014MNRAS.445..899C}.
This latter value is accepted here (assuming neglible difference between Spitzer [3.6] and W1) with the absolute magnitude of the Sun taken to be 5.91~mag AB.

Figure~\ref{fig:2masshist} draws attention to generally recognized properties of galaxies, here within a volume limited sample restricted to $1<d<10$~Mpc and $M_{W1}<-13$~mag.
In the top panel, it is seen that most of the galaxies are dwarfs.
However, in the bottom panel it is seen that overwhelmingly the stellar mass is in the giants ($M_{W1}<-20.5$).

\begin{figure}[!]
\centering
\includegraphics[width=1.00\linewidth]{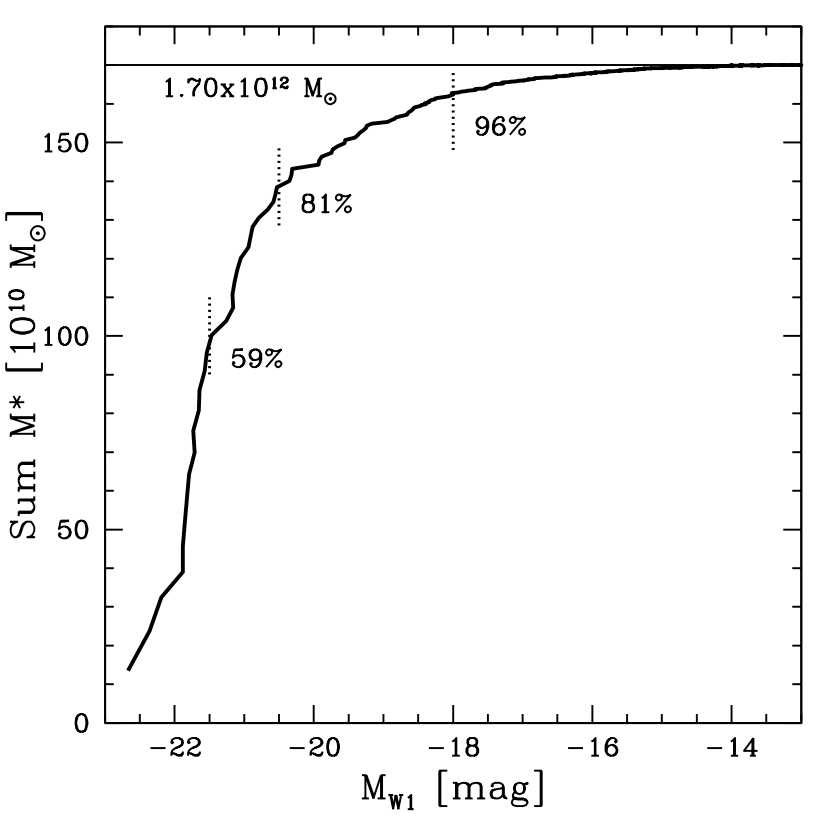}
\caption{Cumulative sum of mass in stars (units $10^{10}~\Msun$) toward fainter $M^{W1}$ magnitudes.  Percentages of total given at $M^{W1}$ values $-21.5$, $-20.5$, and $-18$~mag.}
\label{fig:summass}
\end{figure}

This latter property is emphasized in Figure~\ref{fig:summass}.
Ninety-six percent of the stellar mass is accounted for in galaxies brighter than $M_{W1} = -18$~mag (eg, NGC\,2976 in the M81 group).
Over 80\% of $M^{\ast}$ lies in galaxies brighter than $M_{W1} = -20.5$ (eg, NGC\,5195, the M51 companion).
Almost 60\% lies in the 15 galaxies more luminous than $-21.5$~mag (NGC\,4258 = M106 and brighter).

\section{Summary}
\label{sec:summary}

Photometry of nearby galaxies has been carried out, drawing from the all-sky observations in the mid infrared bands W1 and W2 of the WISE and NEOWISE surveys from space.
The photometry sample includes 567 galaxies with W1 measurements, 558 with W2 values, 515 with W1 lying between one and ten Mpc, 370 of these brighter than $M_{W1} = -13$~mag, and 247 of these with accurate TRGB distances.

Restricting the discussion to the sample with $M_{W1} < -13$ that is expected to be reasonably complete, 80\% of the galaxies lie faintward of $M_{W1} = -18$~mag but 96\% of the stellar mass (inferred from the W1 luminosity) is within the galaxies brighter than this limit. 
There is a remarkable concentration of the stellar mass in this local sample in the half magnitude interval $-21.5$ to $-22$~mag.

This sample has been assembled to further an understanding of the local neighborhood within 10~Mpc: the mass in stars, the disposition into groups and in isolation, the relationships with large scale structure, and bulk motions in response to the distribution of matter, dark and stellar.
These topics will be taken up in a companion paper in preparation.

\begin{acknowledgments}
The authors appreciate the helpful comments by the referee.
Mark Seibert was instrumental in the early development of the GLGA software suite.
This study was supported by NASA grant No.\ 88NSSC18K0424.
\end{acknowledgments}

\newpage
\bibliography{paper}
\bibliographystyle{aasjournal}

\end{document}